\documentclass{article}
\usepackage{arxiv}
\usepackage[utf8]{inputenc} \usepackage[T1]{fontenc}    \usepackage{hyperref}       \usepackage{url}            \usepackage{booktabs}       \usepackage{amsfonts}       \usepackage{nicefrac}       \usepackage{microtype}      \usepackage{lipsum}  \usepackage{listings}
\usepackage{graphicx}
\usepackage{subcaption}
\usepackage{xcolor}
\DeclareUnicodeCharacter{FB01}{fi}
\title{Open Dataset of Phishing and Tor hidden services Screen-Captures}
\author{
  Vincent~Falconieri\\
  CIRCL\\
  Luxembourg\\
}
\begin{document}

\maketitle
\begin{abstract}
Security analysts need to classify, search and correlate numerous images. Automatic classification tools improve the efficiency of such tasks. However, the main resources to develop these tools are \textit{datasets}, which are introduced and provided by the present paper, for the specific cases of visual correlation of phishing and onion websites. CIRCL's Open-Source tools are the sources of these screenshots, which had been manually verified against personal information leaks. Usage examples of these datasets are proposed in the current paper. These researches directions are, however, not the main contribution of the paper. The main contribution is the availability of the two datasets.
\end{abstract}


\section{Introduction}
CERTs such as CIRCL and security teams collect and process content such as images (at large from photos, screenshots of websites or screenshots of sandboxes). 
Datasets become larger - e.g. on average 10000 screenshots of onion domains websites are scrapped each day in \textbf{AIL}\footnote{Analysis Information Leak framework - \href{https://github.com/CIRCL/AIL-framework}{github.com/CIRCL/AIL-framework}}\cite{mokaddemAILDesignImplementation2018}, an analysis tool of information leak - and analysts need to classify, search and correlate through all the images.
Automatic tools can help them in this task. Less research about image matching and image classification seems to have been conducted  exclusively on websites screenshots\cite{sampatCNNTaskClassification}\cite{aburrousPredictingPhishingWebsites2010a}\cite{chenFightingPhishingDiscriminative2009}. However, a classification of this kind of picture needs to be addressed.
Our long-term objective is to build a generic library and services which can at least be easily integrated in \textit{Threat Intelligence tools} such as \textbf{AIL}\cite{mokaddemAILDesignImplementation2018} and \textbf{MISP}\footnote{Malware Information Sharing Platform - \href{https://github.com/MISP/MISP}{github.com/MISP/MISP}}\\\cite{wagnerMISPDesignImplementation2016}. A quick-lookup mechanism for correlation would be necessary and part of this library. This paper includes the release of two datasets to support research effort in this direction.
MISP is an open source software solution tool developed at CIRCL for collecting, storing, distributing and sharing cyber security indicators and threats about cyber security incidents analysis. \\
AIL is also an open source modular framework developed at CIRCL to analyse potential information leaks from unstructured data sources or streams. It can be used, for example, for data leak prevention.
\subsection{Problem Statement}
Image correlation for security event correlation purposes is nowadays mainly manual. No open-source tool provides easy correlation on pictures, without regard to the technology used. Ideally, the extraction of links or correlation between these images could be fully automated. Even partial automation would reduce the burden of this task on security teams. The datasets are part of the foundation needed to construct such tool.\\
\textbf{The main contribution of this paper are two human-verified and human-classified free and open datasets. First dataset is composed of 400+ phishing websites screenshots. Second dataset is composed of 37000+ onion domain website screenshots. Both datasets are available at \href{https://www.circl.lu/opendata/datasets/circl-phishing-dataset-01/}{\textit{https://www.circl.lu/opendata/datasets/circl-phishing-dataset-01}}
 and \href{https://www.circl.lu/opendata/datasets/circl-ail-dataset-01/}{\textit{https://www.circl.lu/opendata/datasets/circl-ail-dataset-01}}
}

This paper presents other contributions : 
\begin{itemize}
\item Human-classification of the datasets. Two classifications and a graph view of these datasets are provided, usable for Image-Matching and Image Classification purposes;
\item A glimpse at research results for visual clustering of phishing websites.
\end{itemize}
Each contribution is described in the following.
This paper is also strongly associated with other contributions not described in the following pages, which are \textit{Carl-Hauser}\footnote{A free and open-source automated benchmarking framework for Image-Matching algorithms review - \href{https://github.com/CIRCL/carl-hauser}{github.com/CIRCL/carl-hauser}} and \textit{Douglas-Quaid}\footnote{A free and open-source library for Image-Matching - \href{https://github.com/CIRCL/douglas-quaid}{github.com/CIRCL/douglas-quaid}}. {\color{red}\textbf{RELATED PAPER}}
\section{Datasets}
\subsection{Source}
Different tools collected the datasets presented by the paper. The list of screenshots' data sources are: 
\begin{itemize}
\item A subset of public security events of proven or potential phishing cases, having attached screenshots, from \textbf{MISP} and from \textbf{URLAbuse}\footnote{Public CIRCL service to review security of a given URL - \href{https://circl.lu/services/urlabuse/}{circl.lu/services/urlabuse/}};
\item A subset of onion domain websites scrapped by \textbf{AIL};
\end{itemize}
At a later date, we could extend these datasets. Complementary data sources may be used, as for example: 
\begin{itemize}
\item Images extracted from DOM. These pictures allow a more particular matching.
\item Screenshots from sand-boxed malware analysis.
\item A subset of analyzed websites by \textbf{Lookyloo}\footnote{A web interface tool allowing to scrape a website and analyze it as well as the tree of called domains - \href{https://github.com/CIRCL/lookyloo}{github.com/CIRCL/lookyloo}}
\end{itemize}
\subsection{Details}
First dataset is named \textbf{circl-phishing-dataset-01} and is composed of phishing websites. Around 460 pictures are in this dataset to date. Three files are provided along with the dataset : one label classification (DataTurks direct output), a second label classification (VisJS transformed output), and a graph-based classification (VisJS direct output).\\
Second dataset is named \textbf{circl-ail-dataset-01} and is composed of AIL's scraped onion websites. Around 37500 pictures are in this dataset to date. Only one label classification (DataTurks direct output) is provided along with the dataset. This classification is partial to date and will be improved and updated as soon as classification operations had been achieved.

\begin{figure}[h!]
  \centering
  \begin{subfigure}[b]{0.49\linewidth}
    \includegraphics[width=\linewidth]{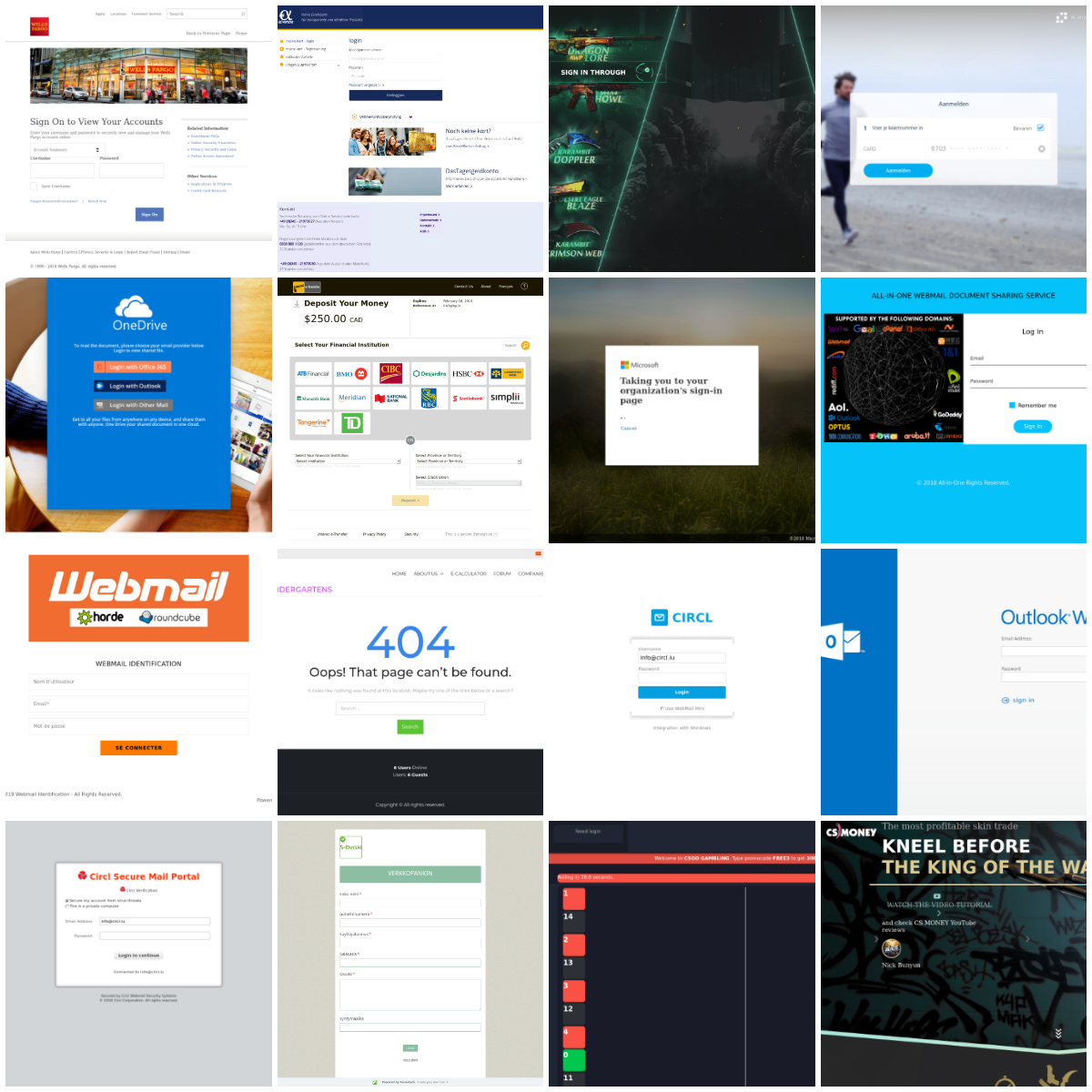}
    \caption{Phishing dataset overview}
  \end{subfigure}
  \begin{subfigure}[b]{0.49\linewidth}
    \includegraphics[width=\linewidth]{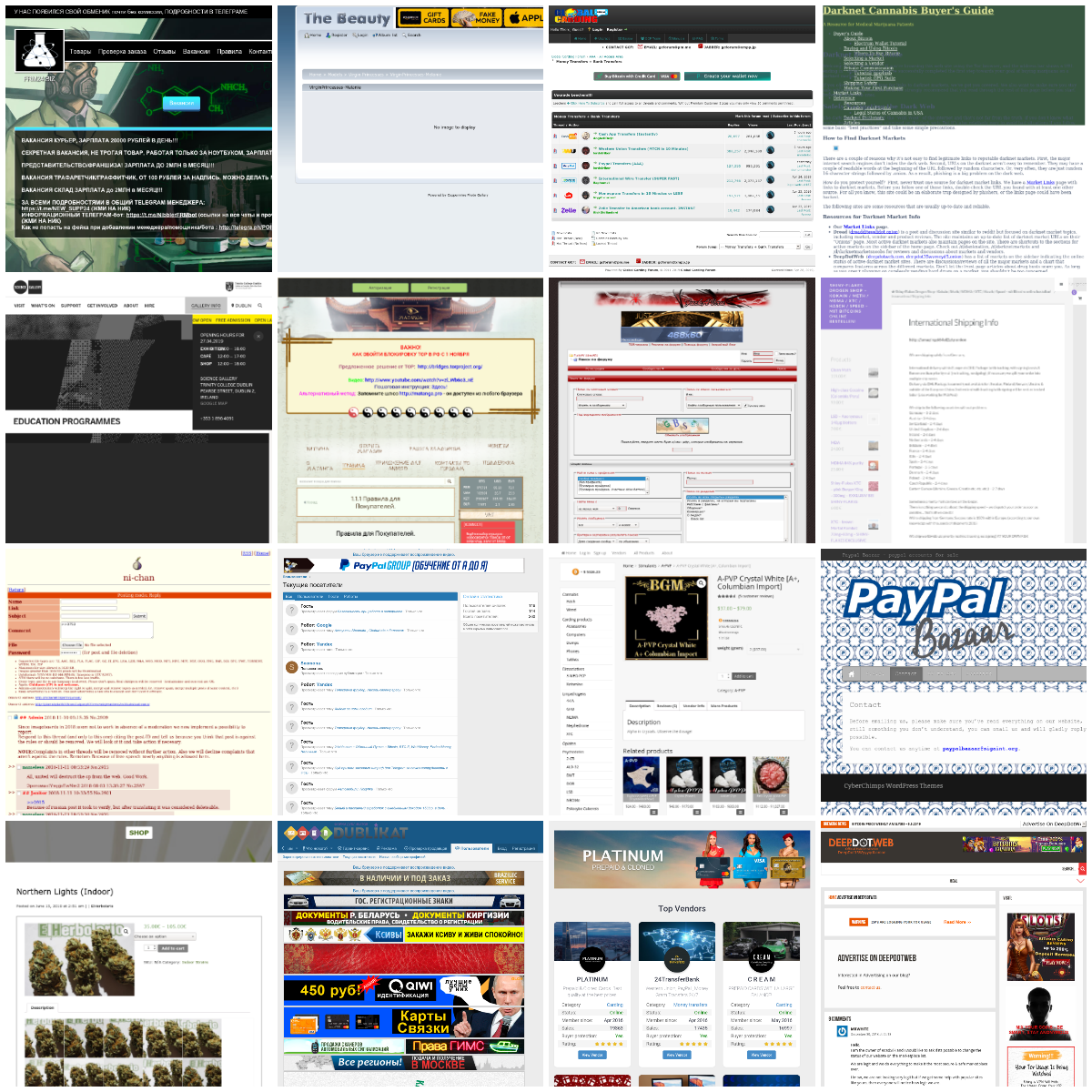}
    \caption{AIL dataset overview}
  \end{subfigure}
  \caption{Dataset's samples}
  \label{fig:datasetsextract}
\end{figure}
\begin{figure}[h!]
  \centering
  \begin{subfigure}[b]{0.37\linewidth}
    \includegraphics[width=\linewidth]{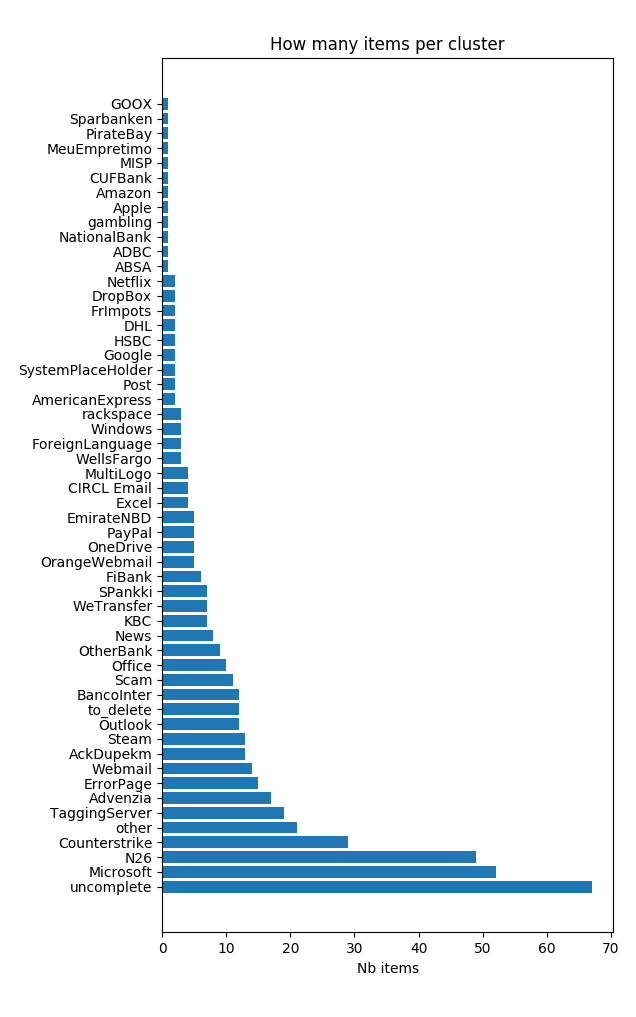}
    \caption{Phishing dataset label frequency}
  \end{subfigure}
  \begin{subfigure}[b]{0.6\linewidth}
    \includegraphics[width=\linewidth]{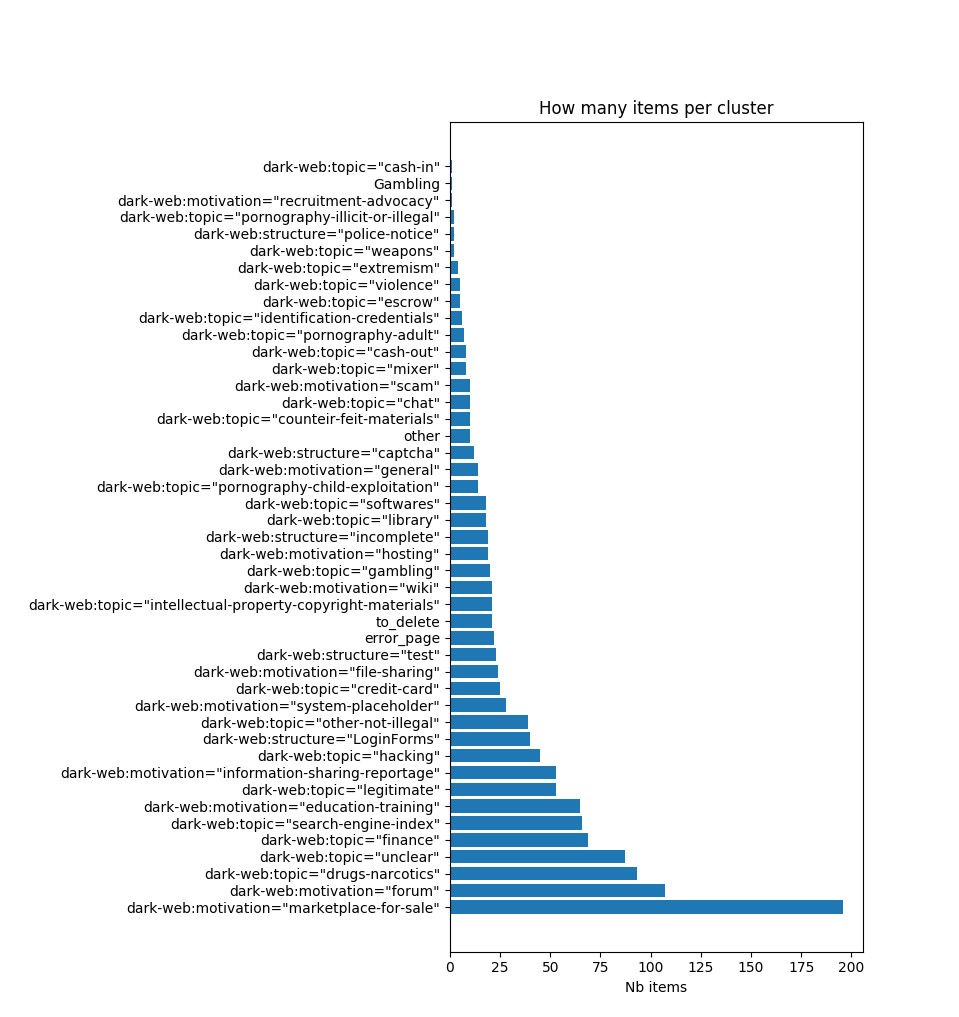}
    \caption{AIL dataset label frequency (on 800 sampled pictures)}
  \end{subfigure}
  \caption{Label frequency per dataset}
  \label{fig:painted}
\end{figure}

\begin{figure}[h!]
  \centering
  \begin{subfigure}[b]{\linewidth}
    \includegraphics[width=\linewidth]{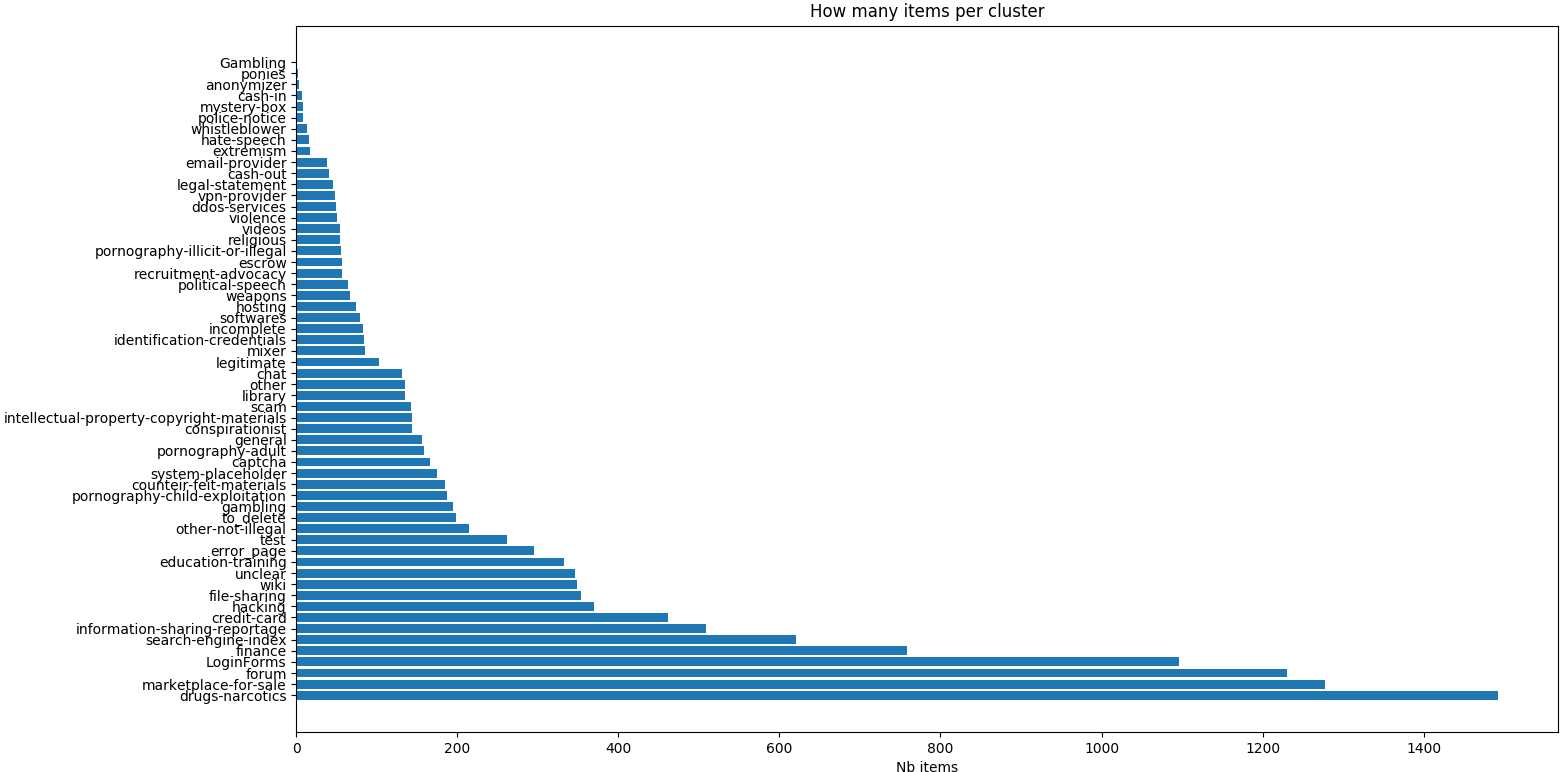}
    \caption{AIL dataset label frequency (on 9500 sampled pictures)}
  \end{subfigure}
  \begin{subfigure}[b]{\linewidth}
    \includegraphics[width=\linewidth]{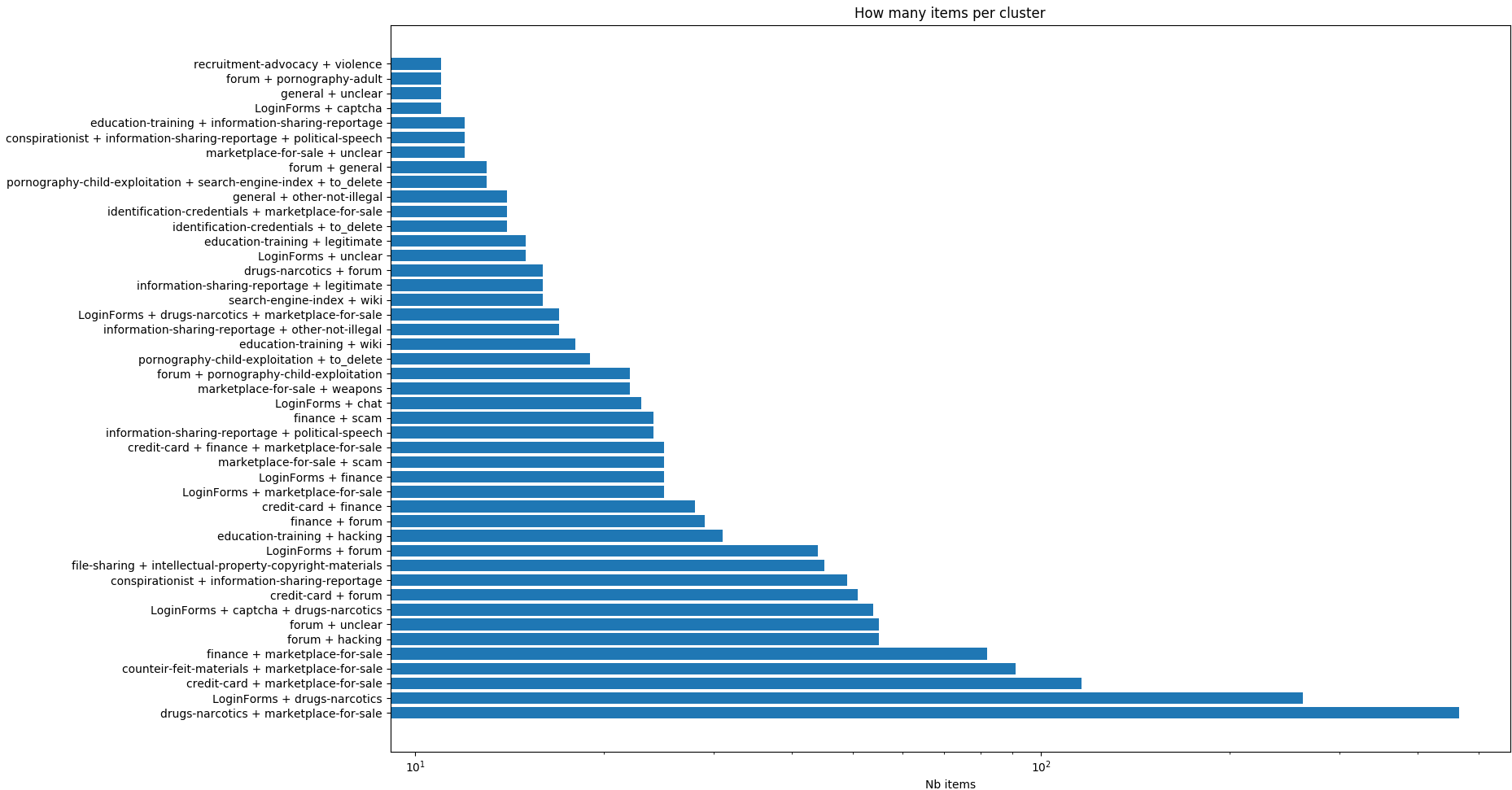}
    \caption{AIL dataset set of labels frequency (on 9500 sampled pictures). Log scale.}
  \end{subfigure}
  \caption{Label frequency for AIL dataset}
\end{figure}

Each picture's content of each dataset was hashed to "humanly readable" name to allow a unified and readable reference system for image's naming convention. This had been performed with a slightly modified version of \textbf{Codenamize.}\footnote{Consistent easier-to-remember codename generator - \href{https://github.com/jjmontesl/codenamize}{github.com/jjmontesl/codenamize}}. The bytes content of each file is hashed and mapped to a list of words, from a dictionary.
Collision was handled by keeping track of which name has already been generated, and temporary adding bytes to each colliding file. However collision was still rare. A human-readable hash of 3 adjectives (without a maximum number of characters) can generate up to 2 trillion combinations, which is far sufficient to handle even 40 000 pictures without common collision occurrence. Collisions were, however, easily met in case of similar pictures (typically, all white or all dark pictures) but then, their name can be swapped without incidence on the meaning of the dataset.
Each dataset is provided as a compressed archive, containing a folder of pictures as well as a reference JSON, containing a mapping from filenames to MD5, SHA1, SHA256 of each picture. This allows an easy retrieval of which picture is which in case picture names need to be modified. Please check file format extract at the section \ref{pictures-hashes}, page \pageref{pictures-hashes}.
\vfill

\clearpage
\pagebreak
\subsection{Classification Methodology}
We manually reviewed datasets, picture by picture. We used a private instance of \textbf{Dataturks}\footnote{OpenSource Data Annotation tool for teams - \href{https://github.com/DataTurks/DataTurks}{github.com/DataTurks/DataTurks}} to perform classification and review the datasets. We removed datasets pictures which were identified as containing personal information such as sensitive e-mail address clearly displayed on screenshots, ... We also manually removed pictures which were identified as containing harmful content, such as violent, offensive, obscene or equivalent undesirable pictures which may shock anyone. \\
We make reasonable effort not to display anything in the dataset which may specifically identify an individual. This dataset is provided for research purposes. \textbf{We stay available for any request.} Please refer to contact information in conclusion of this paper page \pageref{contactinfo}. Please note that each website behind each screenshot can be freely accessed by one with relevant means.
\subsection{Use}
These datasets can be used to create classifiers, which then can be used to automate processes. Few examples of application : 
\begin{itemize}
\item Automatically classify phishing websites : matching screenshots from phishing-like websites to know legitimate websites;
\item Correlate object on pictures from crawled websites, mainly with screenshots of hidden services (AIL use cases);
\item Correlate website screenshots to cluster phishing-like websites together, to keep track of domain-name changes for example (Lookyloo usecase).
\item Isolate and characterise outliers
\item Extracting statistics about crawled websites (per theme, per type, per content, per access allowance ...)
\end{itemize}
\vfill
\pagebreak
\section{Format}
\subsection{Phishing dataset}
\subsubsection{Classification labels}
Labels are used to classify each picture in one or more clusters. Labels are different depending on the dataset and the tool used to classify it.
The phishing dataset was clustered twice : once with Dataturks (Collaborative classic labelling tools), and once with VisJS-Classificator (Collaborative graph-based labelling tools). Distinct labels were used in both tools, as labels were created at classification time. \\

\textbf{These labels were used with Dataturks :}\\
Microsoft, KBC, Advenzia, ErrorPage, WellsFargo, Office, Outlook, Excel, Counterstrike, Steam, other, Scam, uncomplete, TaggingServer, CUFBank, CIRCL, e-mail, AmericanExpress, AckDupekm, BancoInter, N26, HSBC, OneDrive, WeTransfer, DropBox, SystemPlaceHolder, EmirateNBD, Webmail, MultiLogo, PayPal, SPankki, DHL, News, FrImpots, Netflix, FiBank, Sparbanken, OrangeWebmail, Google, ForeignLanguage, MISP, rackspace, MeuEmpretimo, Windows, NationalBank, PirateBay, ADBC, ABSA, GOOX, Amazon, WashingtonTrustBank, OtherBank, gambling, CIBC, Apple, Post \\

\textbf{These labels were used with VisJS :} \\
BancoInter, Advanzia, N26, WeTransfer, microsoft, KBC, multilogo, OrangeWebmail, Netflix, Steam, CounterStrike, Outlook, OneDrive, unloaded, TaggingServer, CirclWebServer, ackouperm, dhl, fibank, paypal, EmiratesNBD, google, SPankki, ForeignLanguage, Office, Windows, Oldcircl, rackspace, DropBox, CarreBlue, WellsFargo, Post, FrImpots, news, AmericanExpress, bitcoin, android, errorMessage, uncomplete, hsbc, formular, misc 
\subsubsection{File structure}
\begin{figure}[h!]
  \centering
  \includegraphics[width=0.8\linewidth]{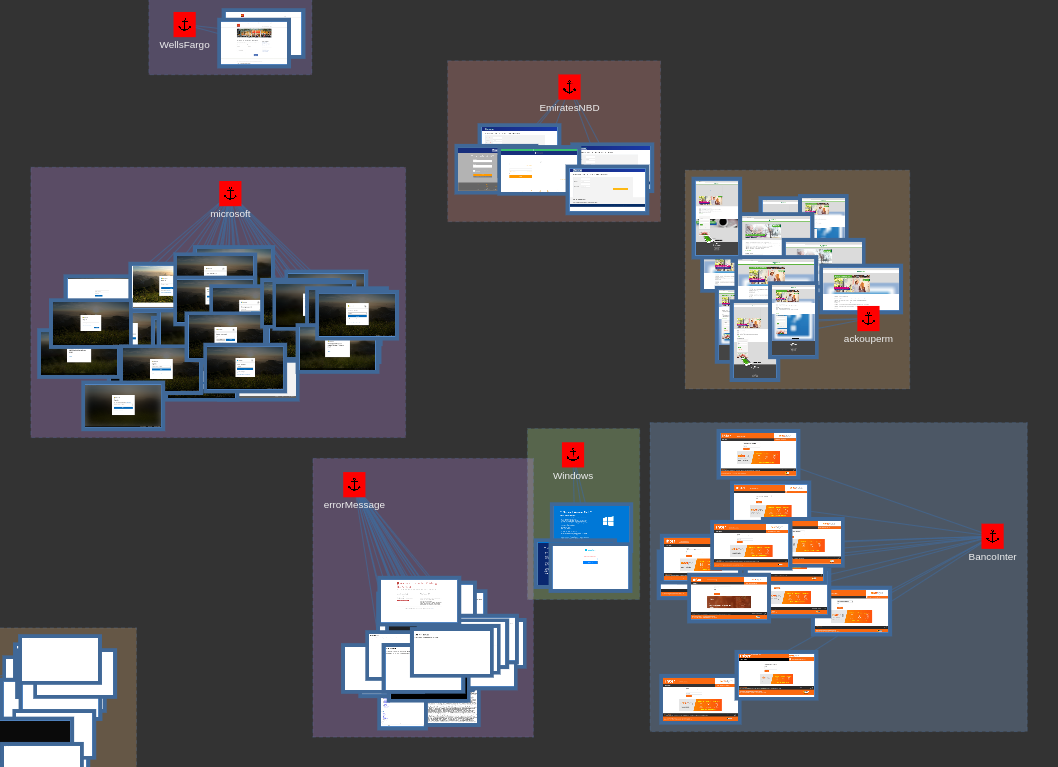}
  \caption{Few examples of clusters found with VisJS Classificator interface.}
  \label{fig:biggestcluster}
\end{figure}

Dataturks tool outputs a list of filename along with their labels.
Please refer to section \ref{dataturks-labels}, page \pageref{dataturks-labels} if you want a technical overview of this file format.
Clustering file formats for VisJS tool is a mapping from cluster name to the list of pictures which belongs to this cluster.
Please refer to section \ref{visjs-labels}, page \pageref{visjs-labels} if you want a technical overview of this file format.
A graph export of VisJS is provided too, which can directly be loaded in VisJS Classificator to get the same display as shown in the picture \ref{fig:biggestcluster}.
\subsection{AIL dark web dataset}
\subsubsection{Classification labels}
Pictures are labelled following MISP "dark-web" taxonomy\footnote{In taxonomies used by MISP and other information threat sharing tools - \href{https://github.com/MISP/misp-taxonomies}{github.com/MISP/misp-taxonomies}}\cite{MISPTaxonomiesClassification}. 
Labels are expressed as triplets 'namespace:predicate=value'.  \\
For example 'dark-web:topic="hacking"' is one label of this taxonomy.
Two labels were added : "error page" and "other" who does not specifically belong to the dark web, but could cover any other case met in the dataset. 
For a complete list of labels used, please refer to section \ref{misp-darkweb-taxonomy}, page \pageref{misp-darkweb-taxonomy}.
\subsubsection{File structure}
Dataturks tool outputs a list of filename along with their labels, to which they belong.
Please refer to section \ref{dataturks-labels}, page \pageref{dataturks-labels} if you want a technical overview of this file format.
\vfill
\pagebreak
\section{Usage example - Automatic Image Matching}
\begin{figure}[h!]
  \centering
  \begin{subfigure}[b]{\textwidth}
    \includegraphics[width=\textwidth]{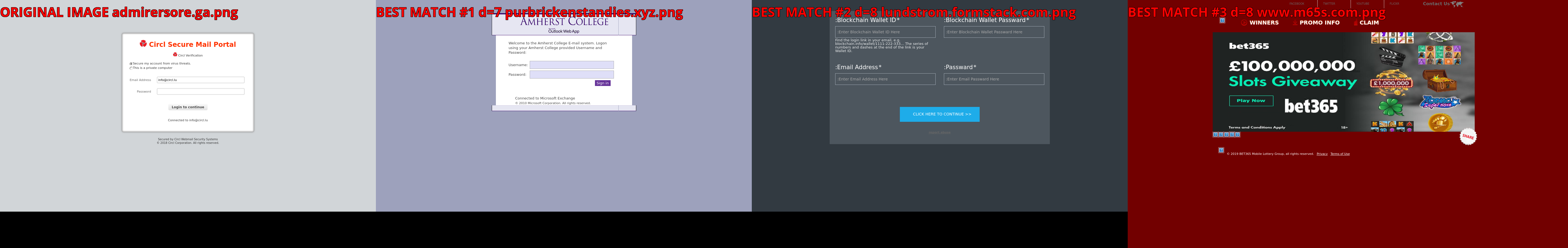} 
    \caption{Structural matching}
    \label{fig:structuralform}
  \end{subfigure}
  \begin{subfigure}[b]{\textwidth}
    \includegraphics[width=\textwidth]{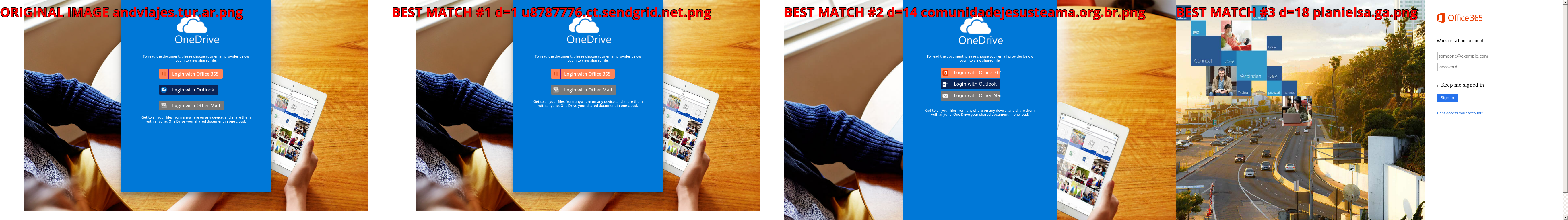} 
    \caption{Theme matching}
    \label{fig:structualmatch}
  \end{subfigure}
  \caption{Usage examples of Phishing dataset to train an Image-Matching tool. Most left-hand picture is requesting pictures, right pictures are best matched pictures. Small differences are in similar pictures, which are a challenge and a common noise that algorithms have to deal with, in a phishing visual-based clustering context.}
  \label{fig:structural}
\end{figure}

This section is a glimpse of what can be achieved with such dataset. Results presented in the following are only preliminary results of \textit{Carl-Hauser}\footnote{A free and open-source automated benchmarking framework for Image-Matching algorithms review - \href{https://github.com/CIRCL/carl-hauser}{github.com/CIRCL/carl-hauser}} operated on the previously presented phishing dataset.
Structures present on pictures can be easily detected with fuzzy-hash approaches, as Figure \ref{fig:structuralform}. Content and themes can be matched with some confidence with fuzzy hashes or feature point approaches, as Figure \ref{fig:structualmatch}.

\section{Future work}
This leads to a list of future possible developments : 
\begin{itemize}
\item Extending provided datasets to support research effort
\item Improve classification provided along with AIL datasets, which need time and human labour
\end{itemize}
\section{Conclusion}
\subsection{Summary}
Phishing datasets and AIL datasets are available for research purposes at \href{https://www.circl.lu/opendata/datasets/circl-phishing-dataset-01/}{\textit{https://www.circl.lu/opendata/datasets/circl-phishing-dataset-01}} and \href{https://www.circl.lu/opendata/datasets/circl-ail-dataset-01/}{\textit{https://www.circl.lu/opendata/datasets/circl-ail-dataset-01}}.
Ground truth files are provided with these datasets at the same address. The datasets and their ground truth file may evolve and be updated.
This research paper proposes that even partial automation of screenshots classification would reduce the burden on security teams, and that the data we provide is a step further in this direction. \\
\subsection{Contact information}
\label{contactinfo}
If you have a complaint related to the dataset or the processing over it, please contact us. We aim to be transparent, not only about how we process but also about rights that are linked to such information and processing. 
You can contact us at \href{https://circl.lu/contact/}{circl.lu/contact/} for requests about the dataset itself, regarding elements of the dataset, or extension requests.
You can contact us at the same address or on \href{https://github.com/CIRCL/carl-hauser}{github}  for feedback about the benchmarking framework, methodology or relevant ideas/inquiries.

\bibliographystyle{paper-ressources/IEEEtran}
\bibliography{./carl-hauser.bib}
\vfill
\pagebreak
\section{Annexes}
\subsection{File format}
\subsubsection{Pictures Hashes}
Each picture has a human-readable name, and is referenced in a JSON file mapping names to MD5/SHA1/SHA256 hashes. The following JSON presents an extract of this mapping available along with both datasets.
\label{pictures-hashes}
\begin{lstlisting}[frame=single,caption=Name to hashes reference file,label=hashes-list]
{
    "abashed-careless-ordinary-crew.png": {
        "md5": "99bbcf5a13c81011fb74cc479c8576a2",
        "sha1": "3e3273eae1f0452f58f63b62f98a0b9a4bd06d68",
        "sha256": "e9beab21d7afbc70ff72abb3c2(...)78ce47837a63ed9"
    },
    "ablaze-jazzy-tangy-file.png": {
        "md5": "b47ab6a326baaae558ed04c9eabb04a0",
        "sha1": "167d9860e301ff99b19500e393eb0e900bfa731c",
        "sha256": "ca4c287aec64c58804826b6d00(...)a92636e981f7eda"
    }, (...)
\end{lstlisting}
\subsubsection{Custom VisJS-Classificator Graph}
CIRCL-Phishing-dataset-01 was labelled with VISJS-Classificator, and the following JSON extract presents the direct graph output of VISJS-Classificator, available at \href{https://www.circl.lu/opendata/datasets/circl-phishing-dataset-01/}{\textit{https://www.circl.lu/opendata/datasets/circl-phishing-dataset-01}} along with pictures.
\label{custom-graph}
\begin{lstlisting}[frame=single,caption=VisJS-Classificator output format,label=graph-struct]
  (...) 
  "clusters": [{
      "id": "a5e1baa2-aead-4164-9205-63f26f656d6f",
      "image": "anchor.png",
      "label": "BancoInter",
      "shape": "image",
      "group": "anchor",
      "members": [
        20,
        (... more members ...)
      ]
    }, (... more clusters ...)],
  "nodes": [{
      "id": 0,
      "image": "abashed-careless-ordinary-crew.png",
      "shape": "image"
    }, (... more nodes ...)
    {
      "id": 456,
      "image": "zonked-silent-snobbish-review.png",
      "shape": "image"
    }],
  "edges": [{
      "to": "a5e1baa2-aead-4164-9205-63f26f656d6f",
      "from": 20
    }, (... more edges / links ...)]
\end{lstlisting}
\vfill
\pagebreak
\subsubsection{Data-Turk labels}
CIRCL-Phishing-dataset-01 was labelled with DataTurks, and the following JSON extract presents the resulting data structure available at \href{https://www.circl.lu/opendata/datasets/circl-phishing-dataset-01/}{\textit{https://www.circl.lu/opendata/datasets/circl-phishing-dataset-01}} along with pictures.
\label{dataturks-labels}
\begin{lstlisting}[frame=single,caption=Dataturks clustering format,label=dataturks-list]
 (...),
    {
        "picture": "tricky-sturdy-impossible-sweet.png",
        "labels": [
            "News"
        ]
    },
    {
        "picture": "old-wet-evasive-influence.png",
        "labels": [
            "N26"
        ]
    }, (...)
\end{lstlisting}
\subsubsection{VISJS labels}
CIRCL-Phishing-dataset-01 was labelled with VISJS-Classificator, and the following JSON extract presents the resulting data structure available at \href{https://www.circl.lu/opendata/datasets/circl-phishing-dataset-01/}{\textit{https://www.circl.lu/opendata/datasets/circl-phishing-dataset-01}} along with pictures.
\label{visjs-labels}
\begin{lstlisting}[frame=single,caption=VisJS-Classificator clustering format,label=VISJS-clustering]
 (...)    
    {
        "cluster": "BancoInter",
        "members": [
            "ambitious-curious-picayune-shock.png",
            "awful-nasty-heartbreaking-appointment.png",
            "charming-defeated-voracious-beat.png",
            "devilish-unwritten-fast-wake.png",
            "greasy-heartbreaking-married-race.png",
            "learned-selfish-elfin-growth.png",
            "magnificent-demonic-faint-recommendation.png",
            "majestic-labored-better-pop.png",
            "muddled-nice-hanging-piano.png",
            "naughty-numberless-wasteful-responsibility.png",
            "robust-drunk-rampant-trouble.png",
            "wrathful-lyrical-alcoholic-county.png"
        ]
    }, (...)
\end{lstlisting}
\vfill
\pagebreak
\subsubsection{MISP dark web taxonomy}
The following triplets extract list is the list of labels used to classify CIRCL-AIL-dataset-01.
Full MISP dark-web taxonomy is available at \href{https://github.com/MISP/misp-taxonomies}{github.com/MISP/misp-taxonomies}
\label{misp-darkweb-taxonomy} 
\begin{lstlisting}[frame=single,caption=MISP dark-web taxonomy,label=misp-taxonomy]
dark-web:topic="drugs-narcotics",
dark-web:topic="extremism",
dark-web:topic="finance",
dark-web:topic="cash-in",
dark-web:topic="cash-out",
dark-web:topic="hacking",
dark-web:topic="identification-credentials",
dark-web:topic="intellectual-property-copyright-materials",
dark-web:topic="pornography-adult",
dark-web:topic="pornography-child-exploitation",
dark-web:topic="pornography-illicit-or-illegal",
dark-web:topic="search-engine-index",
dark-web:topic="unclear",
dark-web:topic="violence",
dark-web:topic="weapons",
dark-web:topic="credit-card",
dark-web:topic="counteir-feit-materials",
dark-web:topic="gambling",
dark-web:topic="library",
dark-web:topic="other-not-illegal",
dark-web:topic="legitimate",
dark-web:topic="chat",
dark-web:topic="mixer",
dark-web:topic="mystery-box",
dark-web:topic="anonymizer",
dark-web:topic="vpn-provider",
dark-web:topic="e-mail-provider",
dark-web:topic="escrow",
dark-web:topic="softwares",
dark-web:motivation="education-training",
dark-web:motivation="file-sharing",
dark-web:motivation="forum",
dark-web:motivation="wiki",
dark-web:motivation="hosting",
dark-web:motivation="general",
dark-web:motivation="information-sharing-reportage",
dark-web:motivation="marketplace-for-sale",
dark-web:motivation="recruitment-advocacy",
dark-web:motivation="system-placeholder",
dark-web:motivation="conspirationist",
dark-web:motivation="scam",
dark-web:motivation="hate-speech",
dark-web:motivation="religious",
dark-web:structure="incomplete",
dark-web:structure="captcha",
dark-web:structure="LoginForms",
dark-web:structure="police-notice",
dark-web:structure="test", 
dark-web:structure="legal-statement",
error_page,
other,
\end{lstlisting}
\end{document}